\documentclass[12pt,preprint]{aastex}

\newcommand\simlt{\lower.5ex\hbox{$\; \buildrel < \over \sim \;$}}
\newcommand\simgt{\lower.5ex\hbox{$\; \buildrel > \over \sim \;$}}
\begin{document}
\title{Baryon Loading of Gamma Ray Bursts by Pick-up Neutrons}
\author{Amir Levinson\altaffilmark{1} and David Eichler\altaffilmark{2}}
\altaffiltext{1}{School of Physics \& Astronomy, Tel Aviv University,
Tel Aviv 69978, Israel; Levinson@wise.tau.ac.il}
\altaffiltext{2}{Physics Department, Ben-Gurion University, Beer-Sheva
84105, Israel; eichler@bgumail.bgu.ac.il}
\begin{abstract}
    It is proposed that the baryons in gamma ray burst (GRB)
fireballs originate as "pick-up" neutrons that leak in sideways
from surrounding baryonic matter and convert to protons in a
collision avalanche. The asymptotic Lorentz factor is estimated,
and, in the absence of collimation, is shown to be angle
dependent. Reasonable agreement is obtained with existing limits
on the GRB baryonic component. The charged decay and collision
products of the neutrons become ultrarelativistic immediately, and
a UHE neutrino burst is produced with an efficiency that can
exceed 0.5. Other signatures may include lithium, beryllium and/or
boron lines in the supernova remnants associated with GRB's and
high polarization of the gamma rays.
\end{abstract}


\section{Introduction}

An outstanding question in GRB fireballs is the fraction of
baryons within them.  It is suspected that this fraction is low,
because they seem to expand with ultra-relativistic Lorentz
factors, $\Gamma \ge 10^2$. While the striking paucity of baryons
could be accounted for by invoking energy release on field lines
where baryons are confined, say by an event horizon or  strong
binding to  strange quark matter, the question would then arise as
to whether the mechanism that enforces baryon purity would be so
effective that there would be none whatsoever in the fireball.

This is possible; the fireball could consist of just pairs, gamma
rays, and low frequency Poynting flux, but then another question
would arise: How do the pairs survive recombination while
expanding from an extremely compact region? Were the fireball
adiabatically expanding, baryon-free and thermal (Paczynski 1986,
Goodman 1986), the $e^+ e^-$ pairs would mostly annihilate at an
internal temperature  of about 15 KeV, corresponding to a radius
of not much more than $\sim 10^9$ cm.  By contrast,
proton-electron pairs would face no such problem, but would raise
the first question: If there are so few, why are there any at all?
[Note that the afterglow from the giant Aug. 27, 1998 flare
(Frail, Kulkarni, and Bloom  1999) from SGR 1900+14 was
sufficiently weak that the outburst seems to have not put much of
its energy into escaping pairs (Eichler 2002), in marked contrast
to long, cosmologically distant GRB's.  This serves as a reminder
that a material fireball that mostly survives the compact regions
of its origin  is not to be taken for granted.]

    A theory of baryon content in GRB fireballs could answer these
questions.  In this letter we attempt to estimate the baryon
content that would arise if the fireball originated entirely on
magnetic field lines that connect to an event horizon (or anything
 equally effective in enforcing total baryon purity). We assume
 that the origin of the baryons is leakage
 in the form of neutrons
crossing from baryon rich field lines  to the baryon free outflow
(Eichler and Levinson, 1999, hereafter EL). The freshly picked-up
neutrons are extremely energetic in the frame of the outflow, and
this creates the possibility of  creating very energetic ($10^9\le
E\ge 10^{15} $eV) pions and neutrinos. The analysis of neutron
pick-up and neutrino production rates and energies is somewhat
similar to that of EL, but here we note a particular instability
that exists in this situation: a collisional avalanche.  This
leads to particularly efficient neutron pick-up and neutrino
production.

Neutrons and neutrinos have also been discussed by other authors.
 Derishev et al. (1999a,b) argued that roughly
equal numbers of neutrons and protons should be present in the
fireball, and studied the consequences for the hydrodynamics and
emission during the prompt and afterglow phases (see also
Beloborodov 2002).  They proposed that decoupling of the neutrons
during the acceleration of the fireball may lead to a burst of
$\sim 10$ GeV neutrinos through inelastic scattering  (see also
Bahcall and Meszaros 2000).  In all the work cited above, the
mechanism of baryon contamination is not addressed. The authors
merely assume that the baryon content of the fireball is fixed by
some mechanism on the smallest scales, prior to its acceleration.
Lemoine (2002) reexamined the conditions under which the neutrons
in the fireball may remain free and concluded that fusion of
neutrons and protons to $^4$He should precede decoupling if the
dynamical timescale exceeds $10^{-4}$ s (cf. Beloborodov 2002).
 He then argued that this would lead to a dramatic suppression
of the density of free neutrons and, consequently, neutrino
emission.  However, he did not discuss the possibility of viscous
heating at the interface between the fireball and its
surroundings.  In what follows we  consider this possibility and
argue that baryonic contamination can occur on much larger scales
($\sim 10^{10}$ cm) via diffusion of neutrons from a baryon rich
wind into the baryon free fireball.

 Additional baryon loading
may arise from mixing, owing to hydrodynamic instabilities at the
interface between the baryon poor jet (BPJ)  and  the baryon rich
outflow (BRW) or the enveloping material of the host star.
However, the degree of mixing may depend strongly on changes in
the magnetic topology, which cannot be easily assessed. Moreover,
it is unlikely that it could load all but the outmost layers of
the BPJ when the hydrodynamic crossing time scale for the dense,
outer material exceeds the proper expansion timescale of the BPJ.

\section{Wind ejection}

    In GRB scenarios that invoke a stellar type progenitor, the
ejection of a baryon poor fireball follows a catastrophic event
that results in the formation of a system consisting of a compact
object surrounded by a hot disk or torus. Here we assume that the
fireball emanates from the immediate vicinity of the compact
object (a black hole, say) essentially devoid of baryons. This
degree of baryon purity is natural if  the fireball is generated
on field lines that thread the black hole, either by neutrino
annihilation or by extraction of the black hole rotational energy
(Levinson and Eichler 1993).

    During the ejection of the fireball, the matter in the central parts is
maintained at temperatures well above 1 MeV, at which emission of
MeV neutrinos is the dominant cooling mechanism. Neutrinos
escaping from the inner parts of the torus heat the surface layers
and drive a powerful, baryon rich wind that propagates at a
sub-relativistic speed, and confines the baryon poor fireball.
Critical point analysis (van Putten \& Levinson 2003, see also
Levinson and Eichler 1993) yields an estimate for the
 baryon rich wind luminosity of
\begin{equation}
L_w\simeq10^{49.5}\beta_{sc}T_{t10}^4A_{13}\ \ {\rm erg\ s^{-1}},
\end{equation}
and for the corresponding mass loss rate of
\begin{equation}
\dot{M}=2L_w/3a_{sc}^2\simeq10^{30}L_{w51}\beta_{sc}^{-2}\ \ {\rm gr\ s^{-1}},
\end{equation}
where $A=10^{13}A_{13}$ cm$^{2}$ is the surface area of the torus,
$T_{t}=10^{10}T_{t10}$ K is the torus temperature
and $a_{sc}=\beta_{sc}c\simeq 10^{10}(M/M_{\sun})^{1/2}(r_c/10^6\ {\rm cm})^{-1/2}$ cm $s^{-1}$
is the sound speed at the wind critical point, with $M$ being the
mass of the compact object, and $r_c$ the radius of the
critical point.  For a temperature of 4 MeV, a wind luminosity of $\sim10^{51}$ erg s$^{-1}$
is expected.

The large optical depth of the expelled wind ($\tau_T\sim 10^8
L_{w51}/r_{10}$) and the extremely short cooling time render the
wind pressure radiation dominated. The determination of the wind's
temperature and density profiles beyond the critical point
requires quantitative treatment of wind acceleration in the
supercritical region. Assuming for simplicity that the wind
velocity $v_w=\beta_w c$ is constant, we obtain a wind temperature
\begin{equation}
T_w\simeq 10^{8.5} L_{w51}^{1/4}\beta_w^{-1/4}(\psi^2-\theta^2)^{-1/4} r_{10}^{-1/2}\ \ {\rm K},
\label{Tw}
\end{equation}
and baryon density
\begin{equation}
n_p=10^{23}L_{w51}/r_{10}^2\beta_w^3(\psi^2-\theta^2)\ \ {\rm cm^{-3}},
\label{np}
\end{equation}
where $\psi$ is the opening angle of the baryon rich wind
(which, as argued below, may be collimated).

If the explosion occurs inside a star, as in the collapsar
scenario for GRB's, both the fireball and the surrounding baryon
rich wind must advance first through the stellar core before they
break out. During this stage they both may be collimated by the
kinetic pressure of the core.  As they advance through the star,
the BPJ and the baryon rich wind each drives a forward shock that
accelerates the stellar material. It is reasonable to assume that
the BPJ and the baryon rich wind have similar isotropic equivalent
luminosities in which case we expect their head velocities to be
roughly the same. The temperature of the shocked wind plasma
(behind the reverse shock) is given approximately by eq.
(\ref{Tw}), upon replacing $\beta_w$ with $\beta_h$. Eventually,
the BPJ breaks through the host star's envelope if there is one.
Otherwise we would not observe the GRB it is supposed to produce.
>From the point of view of neutrino production, however, the
pre-breakout stage may also be important.

\section{The neutron content of the torus wind}
If the torus is a remnant of a degenerate star, then it contains
nearly half its mass in neutrons. At temperatures  above the
dissociation temperature of the nuclei (0.5 - 0.7 MeV) the
neutrons and protons become free. The neutron to proton ratio
$\xi_n$ would then either be the initial value, or,  if the weak
interaction equilibration time
\begin{equation}
t_{weak}\simeq 5\times10^{-2} (T/3 {\rm MeV})^{-5}\ \ s
\end{equation}
is shorter than the outflow time, $t_{exp}$,  the equilibrium
value, $\xi_n\simeq \exp[(m_p-m_n)/T]$, where $m_i$ is the mass of
species $i$ ($i=n,p$). As noted by Derishev et al. (1999b), in the
latter case neutrons will be effectively produced through the
reaction $p+e^{-} \rightarrow n + \nu$, even if the initial torus
composition is predominantly hydrogen. If freeze-out of the weak
interaction occurs at $T>(m_n-m_p)\simeq 1$ MeV, then $\xi_n\sim
1$ is  generally expected.

Above the critical point, the temperature in the wind declines with radius.
Once it drops below about $T_{rec}\simeq77$ keV (depending weakly on density) the free protons
and neutrons will recombine to form deuterium.  The reaction rate for deuterium
formation is $<\sigma_d v>\simeq 5\times 10^{-20}$ cm$^{3}$ s$^{-1}$.   The corresponding
recombination time is then given by
\begin{equation}
t_{rec}\simeq 10^{-3.5}(\psi^2-\theta^2)\beta_w^3r_{10}^2L_{w51}^{-1}\ \ s,
\end{equation}
where eq. (\ref{np}) has been used.  Recombination will be effective at radii
\begin{equation}
r_{10}<10^{3} L_{w51}\beta_w^{-4}(\psi^2-\theta^2)^{-1}
\label{rec-time}
\end{equation}
at which $t_{rec}<<t_{exp}$, provided the temperature there is
below $T_{rec}$. Given $T_{rec}\simeq77$ keV and equation
(\ref{Tw})) we conclude that the radius below which free neutrons
can exist in the wind is
\begin{equation}
r_n\simeq2\times10^{9}L_{w51}^{1/2}(\psi^{2}-\theta^2)^{-1/2}\beta_w^{-1/2}\ \ {\rm cm}.
\label{rn}
\end{equation}
For reasonable wind parameters, $r_n$ should lie in the range
between a few times $10^{9}$ to a few times $10^{11}$ cm.

While we have assumed that the BPJ is ensheathed by a neutron-rich
outflow which we expect exists, we could, for the purposes of the
following discussion,   assume that the BPJ is in direct contact
with the envelope of the host star that collimates it. Neutrons
would be freed up at the interface because the inner wall of the
envelope would be heated by the BPJ.

\section{Neutron pick-up}

The free protons and neutrons in the wind are coupled by nuclear elastic scattering.
At the temperatures
of interest the corresponding rate is
$<\sigma_{el} v>\simeq <\sigma_0 c>=10^{-15}$ cm$^3$ s$^{-1}$,
independent of center of mass energy.

The flux of neutrons diffusing into the BPJ from the interface
separating the BPJ and the baryon rich wind is given by
$J_D(r)=\lambda_{np}v_{is}{\partial n_n}/{\partial x}=
\lambda_{np}v_{is}(n_n/l)$, where
$v_{is}=\beta_{is}c=(kT/m_p)^{1/2}$ the ion thermal speed (not to
be confused with the sound speed of the radiation dominated
plasma), $\lambda_{np}=\beta_{is}/(n_p\sigma_{0})$ is the mean
free path for np collisions, $x$ denotes the cylindrical radius,
and $l$ the gradient length scale (which  may vary with $r$).

 Assuming  that the boundary between the BPJ and baryon rich wind
is very sharp at the BPJ injection radius ($r_0\sim 10^{7}$ cm),
the gradient length scale at some larger radius $r$ is $l\simeq
(\lambda_{np}v_{is}t_{exp})^{1/2}$ with $t_{exp}=r/v_{w}$ being
the wind expansion time. The total number of neutrons diffusing
into the BPJ below some radius $r$ is given by
$N_{diff}(r)\simeq2\pi \theta r^2 t_{exp}J_D(r)= 2\pi \theta r^2 l
n_n$. Combining the above results one finds,
\begin{equation}
N_{diff}(r)=2\pi\theta
\frac{\beta_{is}}{\beta_w^{1/2}}n_n\frac{r^{5/2}}{\sqrt{n_p\sigma_0}}
\simeq10^{50}\xi_n[(\psi/\theta)^2-1]^{-1/2}(\beta_{is}/\beta_w^{2})r_{10}^{3/2}L_{w51}^{1/2},
\label{N_diff}
\end{equation}
where $\xi_n=n_n/n_p$.

 If the density gradient length scale, $l$, at the BPJ baryon
rich wind interface is larger than assumed above, e.g., as a
result of some mixing, then the gradient would be smeared and the
diffusive flux would be reduced, but presumably advection would
replace it and keep the neutron-proton mixture hot. It is hard to
see how mixing could penetrate to the BPJ center without
disrupting the BPJ entirely, so any mixing would be restricted by
this consideration to load only the periphery  of the BPJ.

Now the fraction  of neutrons drifting through the BPJ that decay,
$t_{\rm cross}/\tau_n\simeq 10^{-3.5}\theta
\beta_{is}^{-1}r_{10}$, where $t_{cross}=\theta r/v_{is}$ is the
BPJ crossing time and $\tau_n=900$ s is the neutron lifetime, is
small.  However, each decay liberates a proton that generates more
protons via collisions with the undecayed neutrons. The proton
fraction thus grows exponentially in what we term a collision
avalanche, until becoming comparable to the neutron fraction. To
estimate the growth length of the shower, we note that the density
of target neutrons inside the BPJ (static in the Lab frame) is
$(t_{cross}/t_{exp})(N_{diff}/\pi\theta^2r^3)$, and the optical
depth for a collision of a picked-up baryon with the target
neutrons is
\begin{equation}
\tau_{np}=\sigma_{np}r(t_{cross}/t_{exp})(N_{diff}/\pi\theta^2r^3)=
10^{4.5}\beta_{w}^{1/2}\psi^{-1}r_{10}^{-1/2}L_{w51}^{1/2},
\end{equation}
where a cross section for inelastic pn collisions of
$\sigma_{pn}=40$ mb has been adopted (Hagiwara et al. 2002).
Evidently, the growth length of the shower is much shorter than the injection radius
and it will saturate already at the base of the BPJ.  At this point every neutron diffusing
into the BPJ is picked up via a collision with a fast baryon coming from below.

Combining equations (\ref{rn}) and (\ref{N_diff}), and taking $\beta_{is}=10^{-2}$
(the ion thermal speed at the recombination
temperature) yields the total number of neutrons picked up by the BPJ:
\begin{equation}
N_{cap}=N_{diff}(r=r_n)\simeq 10^{47}\xi_n\theta(\psi^2-\theta^2)^{-5/4}\beta_w^{-11/4}L_{w51}^{5/4}.
\end{equation}
Adopting for illustration $\theta=\psi/2=0.1$, $\xi_n=1$, $\beta_w=0.3$, we obtain
$N_{cap}\simeq 3\times10^{49}L_{51}^{5/4}$.

Now the fact that the avalanche growth is so rapid shows that the
inwardly drifting  neutrons may be converted back to having a 50
percent proton component shortly after crossing into the BPJ, and
will merely line the BPJ outer boundary with a hot viscous
sublayer (just as peripheral  turbulent mixing would do). The
short mean free path also means that relative Lorentz factor
differences across it are likely to be much less than the total
difference. The important assumption is that the inner side is
exposed to contact with the more relaticvistic BPJ and is kept hot
enough to have a free neutron component.

The picture we are led to by our basic model assumptions  is a hot
viscous sublayer with density decreasing inwards, and neutrons and
protons coexisting in roughly equal numbers in the most tenuous
regions.  Below some density, the neutrons stream freely into the
interior of the BPJ.
  Define the free streaming density $n_{fs}$ to be that  at
which the proper hydrodynamic time $r/c\Gamma$ equals the proper
collision time $\Gamma/n<\sigma v>$, i.e.
 \begin{equation}
n_{fs}\equiv \Gamma^2 c/r<\sigma v>.
 \label{nsph}
 \end{equation}
 If the
bulk Lorentz factor $\Gamma$ is determined by the  density of
picked up neutrons then
\begin{equation}
\Gamma_{fs}= L_j/(n_{fs}mc^2\pi \theta ^2 r^2ch) = 26 r_{12}^{-1/3}
L_{j50}^{1/3}\theta^{-2/3}h^{-1/3} \label{gamma}
\end{equation}
where $L_{j}$ is the BPJ luminosity and h is the specific enthalpy
of the fluid in units of $m_pc^2$, and where we have used equation
(\ref{nsph}) in the last part of eq. (\ref{gamma}). The free
streaming density is
\begin{equation}
n_{fs}= (L_j/\pi \theta^2m_pc^3h)^{2/3}r^{-5/3}\sigma^{-1/3}.
\end{equation}

 The number of neutrons per unit time
crossing the free streaming boundary   inward within radius r is
given roughly by
\begin{equation}
 dN_{cr}/dt = \pi \theta  r^2 n_{fs}c/\Gamma = 8\times 10^{49} r_{12}^{2/3}
L_{j50}^{1/3}\theta^{1/3}h^{-1/3}s^{-1}
\end{equation}
where we assume that the random component of the neutron velocity
at the free streaming boundary is close to c. Thus, at
$r_{12}\sim1$, most of the neutrons that diffused into the BPJ at
$r_{10}\le 1$ are already free streaming.

When $n\gg n_{fs}$, h is assumed to be close to unity. In the free
streaming zone, where $\Gamma\gg \Gamma_{fs}$, the neutrons are
subjected to large shear in the BPJ, and  as they move axisward,
find themselves moving relative to the local frame at lab angle
$\chi$ and  nearly backward with a local Lorentz factor
$\gamma'\sim \Gamma \Gamma_{fs}(1-\beta\beta_{fs}cos\chi)$. Thus h
may be estimated as $\gamma'\sim \Gamma
\Gamma_{fs}(1-\beta\beta_{fs}cos\chi)$ in the free streaming zone.
At the free streaming surface, h is very geometry dependent, and
can be between 1 and $\Gamma_{fs}^2$

The local spread of neutron velocities at the free streaming
boundary is of order $1/\Gamma_{fs}$, so if $\theta \ge 1/\Gamma$,
the transverse velocity of an "inwardly" free streaming neutron
may point away from the axis, as long as it does so less than the
average local velocity, and $\chi \le 1/\Gamma_{fs}$. Some
fraction($\sim 1/e$) of the inwardly streaming neutrons will
encounter further collisions, thus loading a "collisional annulus"
of thickness $1/\Gamma_{fs}$ and radius $\theta$, while the rest
may continue further inward
 until they decay. If $\Gamma_{fs}^{2} \Delta t$, where $\Delta t$ is
 the GRB duration, exceeds the neutron lifetime $900s\Gamma_{fs}$,
  then most of the neutrons
 decay within the fireball. Otherwise they decay  behind it and
 leave a baryon pure core ahead.

The fluid parameters in the collisional annulus are similar to
those described by
 Derishev and co-workers, where the  radial acceleration just happens to  proceed
 on a scale comparable to the mean collision time, inducing a
 modestly relativistic
 relative (radial) velocity  between the neutrons and protons. In
 our picture, this apparent numerical coincidence is in fact
 natural for the annulus defined by marginally freely streaming neutrons. The annulus
 occupies $2/\theta \Gamma_{fs}\sim 0.03 r_{12}^{1/3}
L_{j50}^{-1/3}\theta^{-1/3}h^{1/3}$ of the beam solid angle. For
$\theta =0.1$ and $h^{1/3}\sim 6$, this is comparable to the solid
angle of the core, and suggests the possibility that many of the
GRB we see are just these annuli.  Nevertheless, the observational
effects  of a much higher $\Gamma$ core are worth considering,
especially because its elements can overtake  the denser,
peripheral neutron outflow.

It may be that the BPJ is collimated, and that its outer parts -
or at least the inwardly free streaming neutrons -  may eventually
converge at the BPJ axis. The density can increase downstream and
neutrons that have already passed through the free streaming
boundary (according to the formal local definition) may then with
high probability collide further downstream with impact angle
$\chi \gg \Gamma_{fs}^{-1}$in a region of $\Gamma \gg
\Gamma_{fs}$.
(Even in the case of an asymptotically conical BPJ (Levinson and
Eichler 2000), the fact that $\Gamma_{fs}$ decreases with r means
that neutrons free streaming inward from a point $r_1\ll
10^{12}$cm, will graze those free streaming from further upstream
at point $r_2 \le 10^{12}$cm where $r_2\ge r_1$. The impact angle
as viewed in the lab frame is of order  $\Gamma_{fs,2}^{-1}$.)

When $\chi \gg  1/\Gamma_{fs}$, freely axisward streaming neutrons
are "broadsided" by  faster interior plasma at impact angle $\chi
\ge 1/\Gamma_{fs}$, and their optical depth to baryons in the
interior plasma is {\it enhanced} by a factor $\theta^2
\Gamma_{fs}^2$.  Thus, the optical depth presented by axisward
streaming neutrons to the interior plasma is at least of order
unity and neutrino production by picked up particles is efficient.
The energy of the neutrinos released in this avalanche is of order
0.05 of the typical proton energy.

The "top down" nature of pickup suggests that much energy can be
dissipated in extremely energetic collisions.  A high $\Gamma $
flow that is slowed down by neutron pickup can be viewed as
slowing down in stages, such that at each stage $\epsilon(r) dN/dt
\sim L_j$, where $\epsilon(r)$ is the average energy per baryon at
radius r and dN/dt, also a function of r,  is the rate of pickup
within radius r. For each new collisional pickup, about 50 percent
of the original energy is dissipated into pion decay products,
mostly neutrinos. (In each collision between a moving baryon and a
target neutron, many  pions are produced. The leading pion has
about 0.2 of the original baryon energy, and the muon neutrino
from its decay will have about 0.05 of the original, provided that
the impact angle exceeds $1/\Gamma_{fs}$. Thus neutrinos of up to
$\sim 0.05 \epsilon$ will emerge, and they will contain about 5
percent of the jet energy. Another 10 percent or so is in the soft
pions, which have a lab frame Lorentz factor somewhat larger but
of order the center of mass Lorentz factor.) In the crude
approximation that the neutrino losses are a small fraction of the
collision energy, the spectrum of neutrinos is then $dN/dE \propto
E^{-2}$ over a dynamical range that depends on how many optical
depths the avalanche proceeds through; including neutrino losses
leads to an even harder spectrum. Even if neutron pickup is
extremely inefficient, this would still allow efficient neutrino
emission in fewer, higher energy neutrinos. In the extreme
scenario where there is only neutron decay and a single optical
depth for collisions, there are about $6\times 10^{49}c\Delta
t/\tau_n \Gamma_{fs} \sim 10^{48}/\Gamma_{fs}$ neutrons that decay
within the fireball. This suggests an asymptotic  $\Gamma$ of
$10^4$ and a maximum proton energy of
 $10^8m_pc^2$.  Most of this energy will be liberated as neutrinos over
an optical depth of several.  Note that for BPJ proton energies of
order  $10^8m_pc^2$, even the soft pions can decay into $E\gg 1$
TeV neutrinos. Such extremely hard spectra  would be a highly
distinctive signature of the model, and the optical depth crossed
by the avalanche would be manifested  in the neutrino spectrum.

It is also instructive to  view the problem from the point of view
of a free streaming neutron.  It has a better than even chance of
not making another collision once entering the free streaming
zone. However, the rare collision is with extremely relativistic
plasma, and the energy liberated per collision goes as $\Gamma^2$.
Writing the expected energy release per path length as $<dE/ds>=
n<\sigma v>\Gamma^2$, and estimating $\Gamma^2$ as $L/\pi
r^2\theta^2 n$, we notice that $<dE/ds>$ is then independent of
density, and much of the energy release can be in the form of rare
but very energetic collisions.

A  neutrino flux of $10^{52} F_{52}$ ergs  could be detected at
the one count level  with a $ 1 km^2$ neutrino detector at a
distance of $(F_{52}f)^{-1/2}$ Gpc, where f is the beaming factor.
As noted earlier (Eichler and Levinson 1999, Meszaros and Waxman
2001), smothered GRB's could also be detected in (and only in)
neutrinos.

\section{Light Element Production}

Light element production by spallation was discussed by Eichler
and Letaw (1987) in regard to scenarios of UHE particle
acceleration within supernovae ejecta. It was estimated that  each
cosmic ray would produce about 10 $^{10}B$ atoms from interaction
with somewhat heavier nuclei, and that light element abundances
constrain  the average supernova to produce no more than about
$10^{50}$ prompt cosmic rays in the midst of heavy nuclei. For a
shock accelerated spectrum, the number density is dominated by GeV
protons, and this would constrain the energy to less than
$10^{48}$ ergs in cosmic rays. However, this would not apply to a
spectrum of pickup ex-neutrals that is dominated by protons above
1 TeV.  Also, an intense burst of particles on the same surface
would produce "overkill" - i.e. light elements produced by
spallation would then be further degraded by repeated spallation
events. This is significant at a high energy particle fluence
exceeding $10^{26}cm^{-2}$, though Rayleigh-Taylor instabilities
at the head of the jet could lessen the overkill by constantly
providing fresh surface.

Most important, however, is that a GRB-associated supernova is not
typical. It may therefore be worth searching for excess light
element lines in the young supernova remnants associated with
GRB's. Enhancements of two orders of magnitude above the cosmic
abundance could obtain. The original GRB-associated supernova,
SN1998bw, had a remarkably large outflow velocity, c/6, suggesting
that the ejecta may not have been particularly massive, and may
have been preferentially close to the collimated fireball within.

    While energetic protons which could be channeled away from
enveloping  material by electromagnetic forces
 might never interact with the surrounding
 ejecta,  neutrons would easily penetrate the envelope if the ejecta collimate the GRB
fireball, for then the walls of the BPJ curve inward.

\section{Conclusions}

We have worked out the details of an earlier suggestion that
baryon loading of GRB fireballs is accomplished by pick-up
ex-neutrons that crept across magnetic field lines into the path
of the collimated fireball from  the collimating material. For a
GRB lasting, say, 30 seconds ($r_{12}\sim 1$), and having conical
collimation $\theta$, we find that there is a free streaming
annulus at which the flow has a Lorentz factor $\Gamma_{fs}$ of
about $35L_{50}^{1/4}(\theta/0.1)^{-1/2}$. (In this estimate we
assumed that the specific enthalpy h, which is rapidly varying at
this point, is roughly $\Gamma_{fs}$, and note that the result is
only weakly dependent on it.) The thickness of the annulus is
about $1/\Gamma_{fs}$, and $0.6 (\theta/0.1)^{-1/2}$ of the solid
angle within the  cone of opening angle $\theta$ is subtended by
the annulus.  
Well inside the annulus, i.e. when $\theta \gg 1/\Gamma$, the bulk
Lorentz factor may be considerably higher, and the spectrum could
be considerably harder. This suggests the possibility  of
extremely hard GRB's that might yield UHE photons and/or
neutrinos, yet be relatively inconspicuous in soft $\gamma$-rays.
Outside the annulus,  the baryon loading is greater, $\Gamma$ is
considerably lower and this part of the outflow could be
responsible for X-ray flashes (Berezinsky and Prilutsky, 1985),
when the viewing angle happens to coincide with it. Prediction of
a general angular distribution for observed GRB's is frustrated by
our ignorance of the intrinsic distribution for all GRB's, as well
as of the dependence of soft gamma ray efficiency on the amount of
mass loading.

If the GRB is collimated by surrounding material, such as the
envelope of a host star, enough that $\theta \le \Gamma_{fs}$,
then the transverse structure is much closer to being uniform.

 The baryons in the fireball can, of course,
then go on to generate further neutrinos downstream of their point
of origin as in several previous discussions (Eichler 1994,
Paczynski and Xu 1994, Waxman and Bahcall 1997). In this paper, on
the other hand, we have presented a scenario  for baryon loading
whereby UHE neutrinos are a logical consequence.
The neutrinos that result have individual energies  of order
$\Gamma ^2 m_{\pi} \sim  10^{12-15}$ eV, which are  easier  to
detect with large underwater and under-ice neutrino detectors than
those at several GeV, which could result from differential
acceleration of protons and neutrons by fireball pressure. They
would have a very hard spectrum and be a highly distinctive
feature of the pick-up model. Remarkably, the total energy output
in neutrinos can in principle as high as that of the observed
fireball, or even higher.

The ability to catch GRB-associated supernovae at an early stage
should be greatly enhanced by SWIFT, and it may be possible to
search for spallation-induced light element enhancement in the
young supernova ejecta.

The emission of $\gamma$-rays from an annulus just inside an
optically thick wall should give rise to a strongly polarized
reflected component, as noted by EL. A quantitative discussion of
this will be given in a subsequent paper.

We thank E. Derishev and the referee for useful discussions. This
research was supported by the Arnow Chair of Astrophysics at Ben
Gurion University and by an Adler Fellowship awarded by the Israel
Science Foundation.


\begin{thebibliography}{99}
\bibitem[]{} Bahcall, J.N. \& Meszaros, P. 2000, Phys. Rev. Lett., 85, 1362
\bibitem[]{} Beloborodov, A.M. astro-ph/0210522
\bibitem[]{} Berezinsky, V.S. and Prilustsky, O. F. Proc. 19th ICRC, OG 1, 29
\bibitem[]{} Derishev, E.V., Kocharovsky, V.V. \& Kocharovsky, VI.V. 1999a, ApJ, 521, 640
\bibitem[]{} Derishev, E.V., Kocharovsky, V.V. \& Kocharovsky, VI.V. 1999b, A\&A, 345, L51
\bibitem[]{} Eichler, D. Ap. J. Suppl. 99, 877
\bibitem[]{} Eichler, D. 2002,  MNRAS, 335, 4, 883
\bibitem[]{} Eichler, D. \& Levinson, A. 1999, ApJ, 521, L117
\bibitem[]{}  Frail, D., Kulkarni, S. \& Bloom, J.S. 1999, Nature, 398, 127
\bibitem[]{}  Goodman, J. 1986, ApJ, 308, L47
\bibitem[]{} Hagiwara, K. et al. 2002, Phys. Rev. D, 66, 010001
\bibitem[]{}  Lemoine, M. 2002, A\&A, 390, 31
\bibitem[]{} Levinson, A. and Eichler, D. 1993, Ap. J.418, 386
\bibitem[]{}  Meszaros, P., \& Waxman, E. 2001, Phy. Rev. Lett. 87, 1102
\bibitem[]{}  Paczynski, B. 1986, ApJ, 308, L43
\bibitem[]{}  Paczynski, B. and Xu, G. 1994, Ap. J. 427, 708
\bibitem[]{}  Tasevsky, M. (2001) hep-exp 0110084 and references therein
\bibitem[]{} Tan, J., Matzner, C.D. \& McKee, C.F. 2001, ApJ, 551, 946
\bibitem[]{}  van Putten, M.V.P. \& Levinson, A. 2003, ApJ, 584, 937
\bibitem[]{}  Waxman, E. \& Meszaros, P. 2003, ApJ, 584, 390
\bibitem[]{}  Waxman, E. 2001, Nucl. Phys. B (Proc. Suppl.) 91, 494
\bibitem[]{}  Waxman, E. and Bahcall. J. 1997, Phys. Rev. Lett., 78, 2292
\end{thebibliography}
\end{document}